\documentclass[amsmath,amssymb,aps,twocolumn]{revtex4-2}
\usepackage{graphicx}
\usepackage{mathtools}
\usepackage{mathrsfs}

\begin{document}

\title{Transition between the stick and slip states in a simplified model of magnetic friction  }

\author{ Hisato Komatsu }
\affiliation{ Interdisciplinary Graduate School of Engineering Sciences, Kyushu University, Fukuoka 816-8580, Japan }


\email{Present address : Data Science and AI Innovation Research}

\email{Promotion Center, Shiga-university, Shiga 522-0069, Japan}

\email{hisato-komatsu@biwako.shiga-u.ac.jp}

\begin{abstract}
We introduce a simplified model of magnetic friction, and investigate its behavior using both numerical and analytical methods. When resistance coefficient $\gamma$ is large, the movement of the system obeys the thermally activated process. In contrast, when $\gamma$ is sufficiently small, the slip and stick states behave as separate metastable states, and the lattice velocity depends on the probability that the slip state appears. We evaluate the velocities in both cases using several approximations and compare the results with those of numerical simulations.
\end{abstract}

\maketitle

\section{Introduction \label{introduction}}

The microscopic friction mechanism is an important subject of condensed matter physics and engineering\cite{BC06,KHKBC12,PP15,HBPCC94}, and various factors in this phenomenon, such as the lattice vibration and motion of electrons, have been studied\cite{MDK94,DAK98,MK06,PBFMBMV10,KGGMRM94}.
In particular, magnetic friction, the frictional force generated from the magnetic interaction between spin variables, has been extensively studied in recent years\cite{WYKHBW12,CWLSJ16,LG18,RL19}. To understand the mechanism of magnetic friction, many types of theoretical models have been proposed and investigated\cite{KHW08,Hucht09,AHW12,HA12,IPT11,Hilhorst11,LP16,Sugimoto19,FWN08,DD10,MBWN09,MBWN11,MAHW11}.
These models differ from each other in several points such as the definition of dynamics, types of spin variables, and shape of the contact area. Accordingly, important features such as the relation between the frictional force and velocity vary with the choice of model. 

In contrast, there is a well-known empirical law called the Dieterich-Ruina law for normal solid surfaces\cite{Ruina83,Dieterich87,DK94,Scholz98}. This law generally has a complicated form that depends on the hysteresis. However, in the steady state, it can be expressed as the following simple relation:
\begin{equation}
 F = A \log v + B ,
\label{DR}
\end{equation}
where $A$ and $B$ are constants. Note that most of the studies on the magnetic friction ignored the elastic deformation to consider only the influence of the magnetic interaction. Hence, it is difficult to investigate the change of the true contact area generated by the normal force. This is why the dependence on the normal force of the frictional force is not considered in Eq.~(\ref{DR}).
In our previous studies, we proposed models of magnetic friction that seemed to obey Eq.~(\ref{DR}), at least in the steady state\cite{Komatsu19,Komatsu20}. In these models, magnetic structures behaved as a kind of potential barrier that prevented the lattice motion. Hence, we considered that the thermally activated process let the system obey the Dieterich-Ruina law, as in the case of normal solid surfaces\cite{HBPCC94}.
 
\begin{figure}[hbp!]
\begin{center}
\includegraphics[width = 8.0cm]{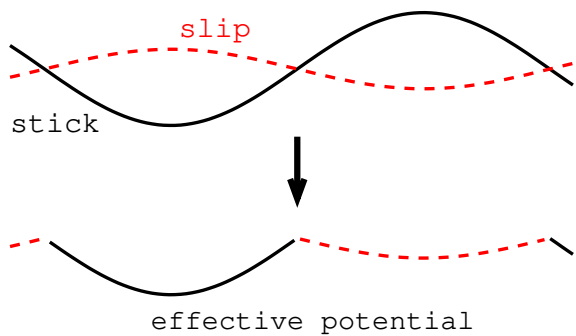} 
\end{center}
\caption{ The effective potential proposed by Ref.~\cite{PAMPSS03}. They proposed that the transition between the stick and slip states occurs when the state with the lowest energy interchanges. }
\label{Persson_eff}
\end{figure}

However, even if the potential barrier exists, the frictional force, $F$, does not always obey the Dieterich-Ruina law. A well-known example is the Prandtl-Tomlinson model, which is one of the earliest theoretical models of friction. This model is composed of one particle moving on a sinusoidal potential pulled by a spring, which represents the contact point between the solid surfaces\cite{PG12,SDG01,Mueser11,PAMPSS03}. If the temperature, $T$, is sufficiently low, this particle is trapped by the potential and cannot move smoothly. This point resembles the phenomenological explanation of the Dieterich-Ruina law. However, as previous studies have explained, frictional force $F$ and velocity $v$ obey the following rule: 
\begin{equation}
F_c - F \propto \left[ - \log \left( \frac{v}{v_c} \right) \right] ^{\frac{2}{3} } ,
\label{PT}
\end{equation}
when $F$ is smaller than a critical value, $F_c$, and the system is trapped by the potential. Here, $v_c$ is the value of $v$ when $F=F_c$. This means that the prevention of the lattice motion by the potential barrier does not always result in Eq.~(\ref{DR}). Note that the systems similar to the Prandtl-Tomlinson model always obey Eq.~(\ref{PT}) when the Taylor expansion of the potential barrier exists, even if the potential is not the sinusoidal one. An example trying to explain the discrepancy between Eqs.~(\ref{DR}) and (\ref{PT}) was proposed by Persson \textit{et al.}\cite{PAMPSS03}. They considered a lubricated system and assumed that the first-order transition of the lubricant played a significant role. Generally, it is known that the friction process consists of two states, the stick state where the lattice is trapped by the potential, and the slip state where it moves smoothly. Persson \textit{et al.} pointed out that the internal structure of the lubricant differs depending on whether the system is in the stick or slip state, and inferred that the system chooses the state with the lowest energy at any given time. Under this assumption, the point where the state with the lowest energy interchanges becomes the singular point of the effective potential, as shown in Fig.~\ref{Persson_eff}. The existence of this point prevents the system from obeying Eq.~(\ref{PT}). Other studies also related the stick-slip motion of a lubricated system to the first-order transition of the lubricant\cite{BDI96,BSF98,XL18}. Magnetic bodies differ from these systems because they do not contain lubricants. However, the phase transition of the magnetization may affect the lattice motion.
To understand the relation between the lattice motion and effective potential produced by the internal degrees of freedom of a solid, including a magnetic structure, a comprehensive investigation using both analytic and numerical approaches is required.

In our previous study, we succeeded in describing the behavior of a kind of infinite-range model in the thermodynamic limit using a mean-field analysis\cite{Komatsu20}. However, we found that the relaxation time of this model diverged with increasing system size, and the actual behavior of the finite-size system in the steady state was different from that of the thermodynamic limit. Hence, it was important to understand the finite-size effect of this infinite-range model. However, an analytical discussion of this effect was difficult because this model had a complicated form. 

Here, we introduce a simplified model to investigate a finite-size system. Specifically, we consider an infinite-range Ising model whose coupling constant $J$ is the function of variable $x$, which represents the shift of the lattice. In our previous models, two magnetic bodies with different order parameters existed\cite{Komatsu19,Komatsu20}. However, for simplicity, our present model contains only one magnetic body with one order parameter, $m$. Although this model is largely simplified, it retains the essence of the magnetic friction, where the magnetic structure prevents the lattice motion. As we will explain later, we calculated the histogram of the state of this system, and found that the distinctions between the stick and slip states could be classified into two types.

In the reminder of this study, we first introduce the model in Sec.~\ref{model}. We then calculate its behavior in Sec.~\ref{calculations}, and finally summarize the study in Sec.~\ref{summary}. Supplementary discussions are also included as appendices~\ref{AppA}, \ref{AppB}, and \ref{AppC}.

\section{Model \label{model} }

We here consider $N$ Ising spin variables $\left\{ \sigma_i \right\} $ interacting with each other by the following Hamiltonian:
\begin{equation}
H = -\frac{J(x) }{N} \sum_{i,j} \sigma _{i} \sigma _{j} = -N J(x) m^2 .
\label{Hamiltonian}
\end{equation}
Here, $x$ is a real value representing the shift of the lattice, $m$ is the magnetization per spin: $m \equiv \sum _i \sigma _i /N $, and $J(x) $ is the periodic function of $x$. For simplicity, we impose a periodic boundary condition with period $2 \pi$ on $x$. Note that we consider a model composed of one magnetic body with one order parameter, $m$, whereas our previous models had two magnetic bodies with different order parameters. If we try to consider the situation that two magnetic bodies interact with each other, we should construct a model containing at least their magnetizations $m_a$ and $m_b$. 
The model containing only one order parameter like Eq.~(\ref{Hamiltonian}) corresponds to the case that the other parameter is fixed as $m_b = \mathrm{const.}$ by the strong internal interaction, for example.
 Furthermore, the ferromagnetic order appears in the case of this model, whereas the antiferromagnetic one appeared in our previous studies. However, even in this simplified model, the magnetization, $m$, behaves as a potential barrier that prevents lattice motion. That is, when $m$ gets larger, the Hamiltonian also has the larger value and trap the lattice motion more easily. Hence, we consider that the essence of the magnetic friction is not lost through this simplification. 
As a concrete form of $J(x)$, two types, A and B, are considered. Type-A is a piecewise linear function given by the following: 
\begin{equation}
 J(x) = J_0 + J_1 \left(1 - \frac{2}{\pi} |x| \right) \ \ \mathrm{if} \ \ -\pi \leq x \leq \pi  , \label{coupling_A1} \\
\end{equation}
\begin{equation}
 J(x + 2 \pi ) = J(x) , \label{coupling_A2}
\end{equation}
and type-B is a sinusoidal function:
\begin{equation}
 J(x) = J_0 + J_1 \cos x .
\label{coupling_B}
\end{equation}
This model coincides with the infinite-range Ising model with the exception that coupling constant $J$ depends on the shift in the system, $x$. In this study, we let $J_0 = J_1 = 1$. In both type-A and B models, the maximum of $J(x)$ is $J(0) = J_0 + J_1=2$. If the lattice was fixed at $x=0$, the critical temperature of the spin variables would be given as $T_c = 2J(0)=4$. In the following, we consider the cases that $T<4$ to investigate the relation between the magnetization and lattice motion.

We introduce the time development of spin variables by the Glauber dynamics, and define one Monte Carlo step (MCS) as the unit time. Namely, the acceptance ratio, $w$, of each updating of the spin variable is defined by the following equation:
\begin{equation}
w \left( \beta \delta E \right) \equiv \frac{1 - \tanh \left( \frac{\beta \delta E}{2} \right)}{2} ,
\label{Glauber}
\end{equation}
where $\delta E$ is the change in energy. In addition, the time development of $x$ under this Hamiltonian and external force $F_{\mathrm{ex} }$ are introduced. Specifically, we let $x$ obey the following overdamped Langevin equation:
\begin{equation}
\gamma N \frac{dx}{dt} = F_{\mathrm{ex} } -\frac{\partial H }{\partial x} + \sqrt{ 2 \gamma NT } R(t) ,
\label{Langevin1}
\end{equation} 
where $R(t)$ is the white Gaussian noise, $\left< R(t) R(t') \right> = \delta (t-t' )$. Here, we introduce coefficient $N$, considering the situation that $N$ spins move simultaneously. Substituting Eq.~(\ref{Hamiltonian}) and letting $f_{\mathrm{ex} } \equiv F_{\mathrm{ex} }/N$, this equation can be transformed as follows:
\begin{eqnarray}
\frac{dx}{dt} & = & \frac{f_{\mathrm{ex} } }{\gamma} -\frac{1}{\gamma N } \frac{\partial H }{\partial x} + \sqrt{ \frac{2T}{\gamma N } } R(t) \nonumber \\
 & = & \frac{f_{\mathrm{ex} }  + J'(x) m ^2 }{\gamma}  + \sqrt{ \frac{2T}{\gamma N } } R(t) .
\label{Langevin2}
\end{eqnarray} 
Considering Eq.~(\ref{Langevin2}), magnetization $m$ prevents the lattice motion when $J'(x) < 0$. In the steady state, the frictional force balances the external force, $f_{\mathrm{ex} } $. Hence, to investigate whether the system obeys Eq.~(\ref{DR}), (\ref{PT}), or other rules, the $v$-$f_{\mathrm{ex} }$ relation should be considered.

In the thermodynamic limit, $N \rightarrow \infty$, the random-force term of Eq.~(\ref{Langevin2}) disappears, and the dynamics of $x$ become deterministic:
\begin{equation}
\frac{dx}{dt} = \frac{f_{\mathrm{ex} }  + J'(x) m ^2 }{\gamma} .
\label{dxdt_inf}
\end{equation} 
As we already mentioned, the Hamiltonian given by Eq.~(\ref{Hamiltonian}) has almost the same form as the normal infinite-range Ising model. Therefore, the time development of $m$ in $N \rightarrow \infty$ is given by the following\cite{SK68}:
\begin{equation}
\frac{d m }{dt} = -m + \tanh \Bigl( 2\beta J(x)m \Bigr) .
\label{dmdt_inf}
\end{equation} 
Note that $m$ coincides with its ensemble average, $\left< m \right>$, in this limit, because the fluctuation of $m$ is $O(1/\sqrt{N} )$. According to the above discussions, the behavior in the thermodynamic limit, $N \rightarrow \infty$, can be described by two deterministic equations (\ref{dxdt_inf}) and (\ref{dmdt_inf}).

\section{Calculations \label{calculations}}

In this section, we investigate the behavior of this system using both numerical simulations and analytic methods. In the simulation, $x$ is updated by the stochastic Heun method with time interval $\delta t = 1/(nN)$, where $n$ is an integer. This means that we update $x$ every $1/n$ step of the Glauber dynamics. In other words, we repeat the updating of $x$ by Eq.~(\ref{Langevin2}) $n$ times after one updating of the spin variables. This process is introduced because the change in $x$ during one step of the Glauber dynamics is large when $\gamma$ is small. We let $n=20$ for $\gamma \leq 0.01$, and $n = 5$ for larger values of $\gamma$. 

\subsection{Behaviors in the thermodynamic limit}

\begin{figure}[hbp!]
\begin{center}
\includegraphics[width = 8.0cm]{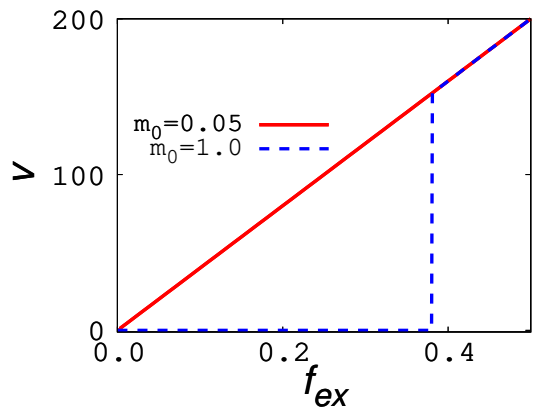} 
\end{center}
\caption{ Relation between (the time average of) velocity $v$ and external force $f_{\mathrm{ex} }$ of the type-A model at $T=3$ and $\gamma=0.0025$. The red solid and blue dashed curves indicate the cases where the initial condition is given as $(x,m) = (0,0.05)$, and $(0,1)$, respectively.  }
\label{vinf_A}
\end{figure}

First, we discuss this system in the thermodynamic limit, $N \rightarrow \infty$. As we explained in Sec.~\ref{model}, the time development of the system is described by Eqs.~(\ref{dxdt_inf}) and (\ref{dmdt_inf}). We calculate (the time average of) velocity $v$ in the steady state by solving these equations using the fourth-order Runge-Kutta method with time interval $\delta t = 10^{-4}$. Fig.~\ref{vinf_A} shows the relation between $v$ and $f_{\mathrm{ex} }$ for the type-A model at $T=3$ and $\gamma=0.0025$. Here, two types of initial conditions are considered, $(x,m) = (0,0.05)$ and $(0,1)$, and the velocity is averaged over $10^3 < t < 4 \times 10^3$. As seen in Fig.~\ref{vinf_A}, there is a range where the velocity in the steady state depends on the initial value.

\begin{figure}[hbp!]
\begin{center}
\includegraphics[width = 8.0cm]{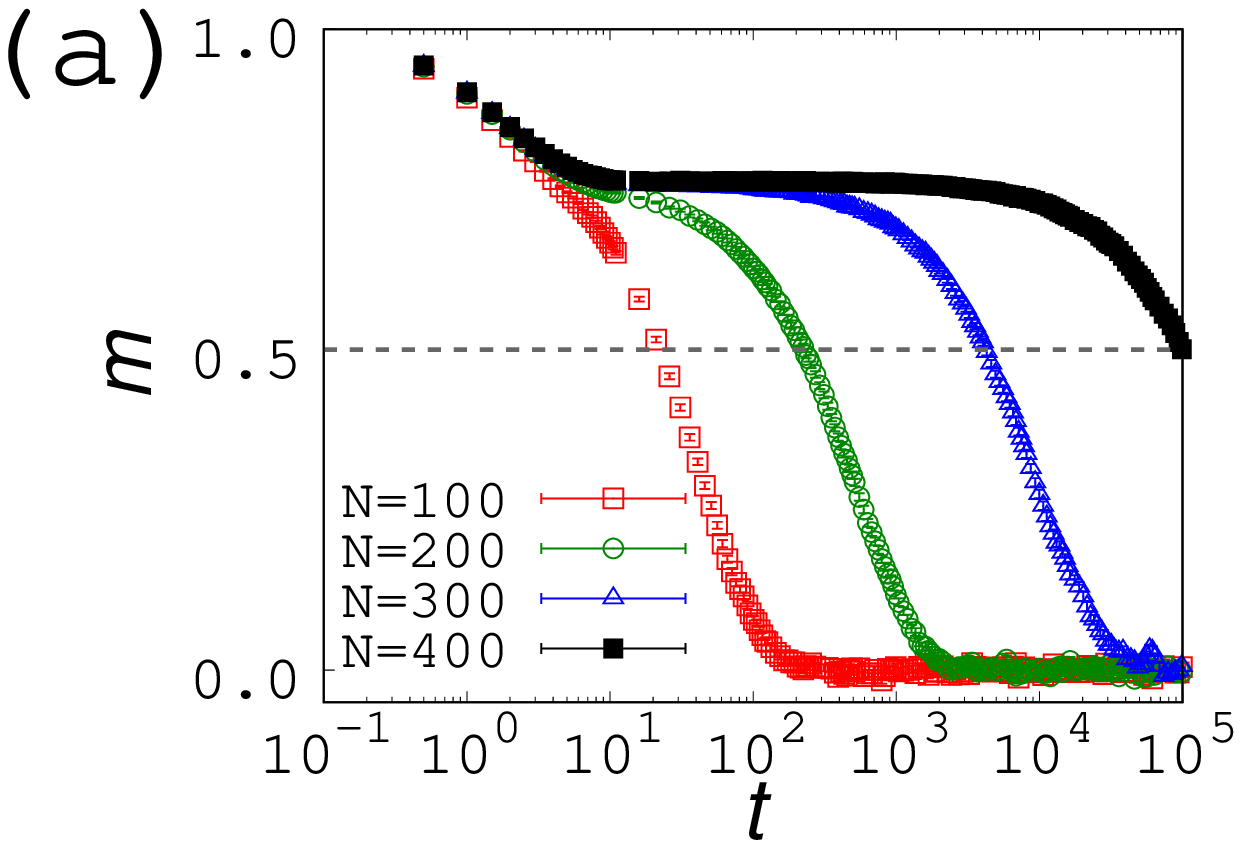} 
\includegraphics[width = 8.0cm]{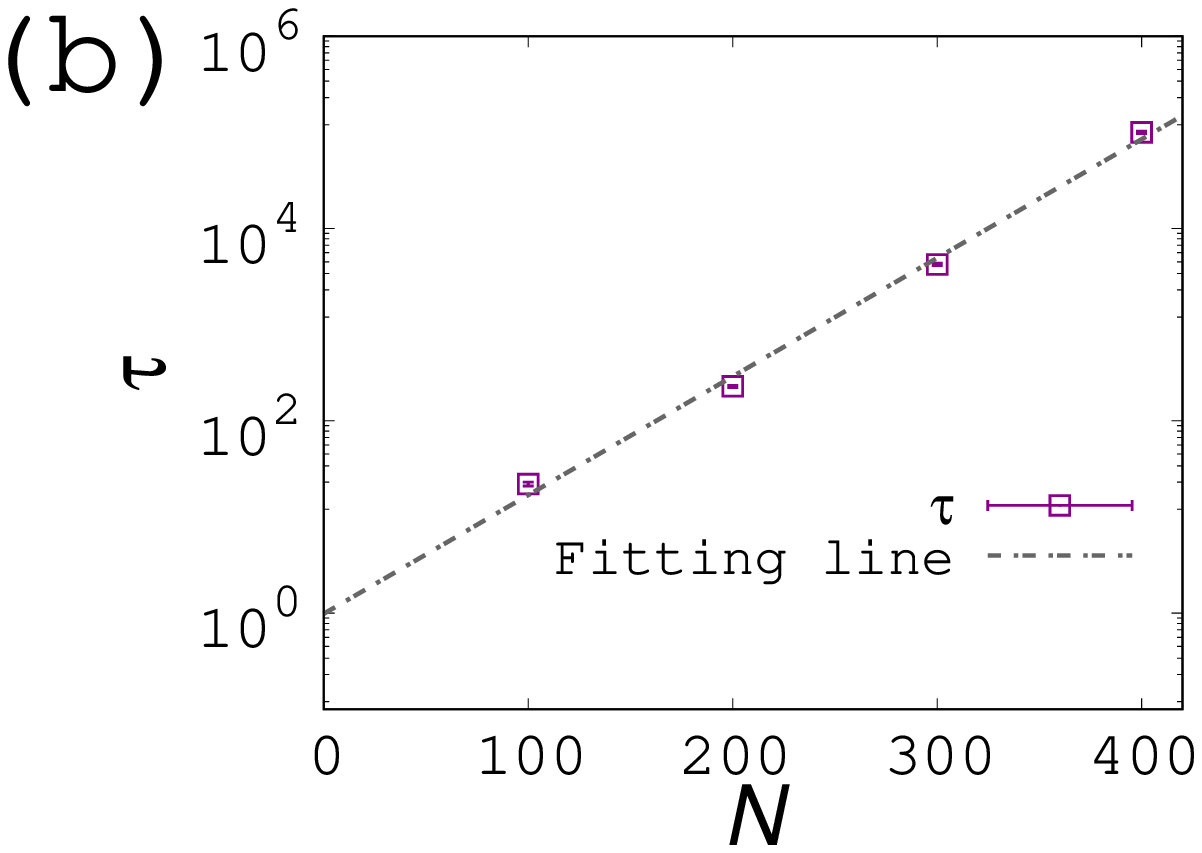}
\end{center}
\caption{ (a)Relaxation of magnetization $m$ of the type-A model at $T=3, \gamma=0.0025$, and $f_{\mathrm{ex} }=0.1$. The red open square, green circular, blue triangular, and black closed square points represent the data with $N=100$,200,300, and 400, respectively. (b)Relaxation time $\tau$ under the same condition as graph (a). Here, $\tau$ is defined as the time when $m$ becomes smaller than the threshold value, $m=0.5$ (the gray dashed line of graph (a).) Data are plotted as the purple square points, and the gray dash-dotted line expresses the fitting line, $\log \tau = 0.0284 N - 0.0141$. }
\label{relax_A}
\end{figure}

We also calculated the time development of the finite-size system using numerical simulation. In this simulation, we chose the type-A model, let $T=3, \gamma=0.0025$, and $f_{\mathrm{ex} } = 0.1$, and started from the perfectly ferromagnetic state with $x=0$. To obtain the data with error bars, each quantity was averaged over 6400 independent trials. The result is shown in Fig.~\ref{relax_A} (a). We also calculated the relaxation time $\tau$ in Fig.~\ref{relax_A} (b). Here, $\tau$ is defined as the time when $m$ becomes smaller than the threshold value, $m=0.5$ (the gray dashed line of graph (a).) As seen in the fitting line of Fig.~\ref{relax_A} (b), $\tau$ behaves as the exponential function of $N$, and diverges rapidly in the thermodynamic limit. Hence, it is natural that the system shows dependence on the initial condition in this limit. The divergence of the relaxation time was also observed in the infinite-range model of our previous study\cite{Komatsu20}, and made the consideration of the finite-size system difficult. Hence, we should investigate our present model, which is much simpler, to understand the behavior of a finite-size system. 
Note that we also performed similar calculations for the type-B model, but omitted the graphs because no qualitative differences between Figs.~\ref{vinf_A} and \ref{relax_A} were found.

\subsection{Behaviors of the finite-size system \label{Histmx} }

We then investigated the behavior of the finite-size system in the steady state, mainly using numerical simulation. In the simulation, we started from the perfectly ferromagnetic state with $x=0$ and a sufficiently weak external force, $f_{\mathrm{ex} }$, and then gradually increased $f_{\mathrm{ex} }$. At each value of $f_{\mathrm{ex} }$, the first $2 \times 10^5$ MCSs were used for the relaxation, and the next $8 \times 10^5$ MCSs were used for the measurement. We mainly investigated the cases where $T=1.5, N=150$ and $T=3, N=300$. The value of $N$ at each $T$ was chosen so that the relaxation time became smaller than the simulation time. To obtain the data with error bars, each quantity was averaged over 64 independent trials.

\begin{figure}[hbp!]
\begin{center}
\includegraphics[width = 8.0cm]{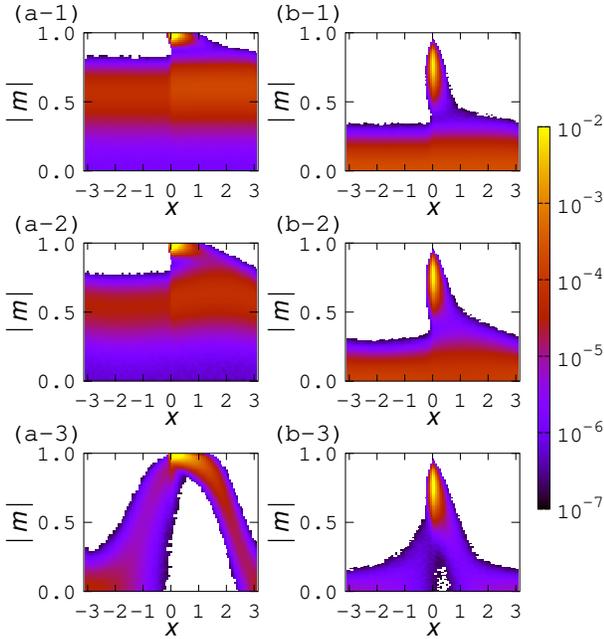} 
\end{center}
\caption{ Histograms of the type-A model at (a)$T=1.5, N=150$ and (b)$T=3,N=300$. In each column, the three graphs have different values of $\gamma$: (a-1)$\gamma=0.0025, f_{\mathrm{ex} }=0.5$, (a-2)$\gamma=0.04, f_{\mathrm{ex} }=0.52$, (a-3)$\gamma=2, f_{\mathrm{ex} }=0.57$, (b-1)$\gamma=0.0025, f_{\mathrm{ex} }=0.1$,  (b-2)$\gamma=0.04, f_{\mathrm{ex} }=0.12$, and (b-3)$\gamma=2, f_{\mathrm{ex} }=0.14$. Note that we chose an appropriate value for  $f_{\mathrm{ex} }$ to clearly see the structure of the histogram at each $\gamma$. }
\label{Histmx_A}
\end{figure}

To investigate the relation between the magnetic structure and lattice motion, we first calculated the histograms of $m$ and $x$. To make these histograms, we divided the interval of $x$ into 100 parts, and did not color the points where the (relative) frequency was lower than a threshold value, $10^{-7}$. Fig.~\ref{Histmx_A} shows several examples of these, in the case of the type-A model. As shown in  Figs.~\ref{Histmx_A} (a-3) and (b-3), when $\gamma$ is large, the value of magnetization $m$ changes continuously with $x$. This means that the lattice motion and weakening of $m$ occur simultaneously. This process is thought to be caused by thermal activation if $f_{\mathrm{ex} }$ is small. Conversely, in a case where $\gamma$ is small, two peaks are observed in the histograms (see Figs.~\ref{Histmx_A} (a-1), (a-2), (b-1), and (b-2).) One of these exists near the fixed point of Eqs.~(\ref{dxdt_inf}) and (\ref{dmdt_inf}), and the other has the form of a band with relatively small $m$. These peaks are thought to indicate the stick and slip states, respectively. This fact means that the stick and slip states are clearly separated as different metastable states. Hence, the change between these states resembles the first-order phase transition, when $\gamma$ is sufficiently small. We also calculated the histograms of the type-B model, but omitted the graph because the qualitative behavior was similar to Fig.~\ref{Histmx_A}. Namely, even in the type-B model, a first-order-like transition was observed in the small-$\gamma$ system, and the thermal activation process was found in the large-$\gamma $ system. The border between these two behaviors are discussed in the appendix \ref{AppA}. Judging from Eq.~(\ref{Langevin2}), the velocity tends to have a large value when $\gamma$ is small. Hence, it is thought that fast lattice motion prevents the slip state from changing into the stick state, and consequently these states are separated from each other. The point that the stick-slip motion is related to the first-order transition resembles the inference proposed by Ref.~\cite{PAMPSS03}. However, the lattice motion of our model (at a small $\gamma$) occurs only in the slip state, while their theory considered that the lattice moved with the stick and slip states switching. Note that the stick and slip states of our model do not interchange smoothly because they are metastable states. 

\subsection{Evaluation of the effective potential under small $\gamma$ \label{evaluation} }

It is difficult to calculate the theoretical probabilities with which the stick and slip states appear because the lattice motion is a non-equilibrium phenomenon. However, when $\gamma$ is extremely small, the change in $x$ is much faster than that in $m$. In this case, it is expected that the effective free energy under a given $m$ could be evaluated by taking a kind of time average over $x$. In this section, we evaluate the probability of the slip state using this effective free energy.

Under the assumption of this section, the time scales of $m$ and $x$ are completely separate. Hence, we should take time average of Eq.~(\ref{dmdt_inf}) to discuss the contribution of $J(x)$ to the time development of $m$:
\begin{equation}
\frac{d m }{dt} = -m + I_1 (m) .
\label{dmdt0}
\end{equation} 
Note that Eq.~(\ref{dmdt_inf}) itself describes the behavior of the thermodynamic limit, as explained in Sec. \ref{model}. However, we use this relation for the following discussions assuming that the effect of the fluctuation of $m$ on the effective free energy is small. In Eq.~(\ref{dmdt0}), $I_1$ is defined as follows:
\begin{eqnarray}
I_1 (m) & \equiv & \int _0 ^{2 \pi} dx \cdot \tanh \bigl( 2\beta J(x) m \bigr) p_m (x) , \label{I1_def}
\end{eqnarray} 
where $p_m (x)$ is the probability distribution of the Brownian motion given by Eq.~(\ref{Langevin2}) under a fixed $m$ in the steady state.

We then evaluate the effective free energy per spin, $a_{\mathrm{eff} } (m)$, using Eq.~(\ref{dmdt0}). To extract the contribution of the entropy, the effective ``internal energy,'' $u_{\mathrm{eff} } (m)$, is defined by the following equation:
\begin{equation}
a_{\mathrm{eff} } (m) = u_{\mathrm{eff} } (m) - Ts(m) ,
\label{Free_slip}
\end{equation} 
where $s(m)$ is the entropy of the system per spin, under a given $m$:
\begin{equation}
s(m) = -\sum _{\eta = \pm 1} \frac{1+ \eta m}{2} \log \left( \frac{1+ \eta m}{2} \right) .
\label{entropy1}
\end{equation} 
To evaluate $u_{\mathrm{eff} } (m)$, we consider the time development of $m$ in a case where the energy of the system per spin is really given as $u_{\mathrm{eff} } (m)$, and compare it with Eq.~(\ref{dmdt0}). 

The probability that one up (down) spin is reversed at each step of the updating, $P_{\mathrm{u \rightarrow d} }$($P_{\mathrm{d \rightarrow u} }$), is given by the following: 
\begin{eqnarray}
P_{\mathrm{u \rightarrow d} } & = & \frac{1+ m }{2} w \Bigl( -2 \beta u' _{\mathrm{eff} } \left( m \right) \Bigr) \label{Pud} \\
P_{\mathrm{d \rightarrow u} } & = & \frac{1- m }{2} w \Bigl( 2 \beta u' _{\mathrm{eff} } \left( m \right) \Bigr) .
\label{Pdu}
\end{eqnarray} 
Here, the change in energy is evaluated based on the fact that reversing the up (down) spin indicates a decrease (increase) in $m$ by $2/N$. Thus, the change in the expectation value, $\left< m \right>$, is expressed as follows:
\begin{eqnarray}
& & \left. \left< m \right> \right|_{t+\frac{1}{N} } - \left. \left< m \right> \right|_{t } \nonumber \\
 & = & \frac{2}{N} \left( P_{\mathrm{d \rightarrow u} } - P_{\mathrm{u \rightarrow d} } \right) \nonumber \\
 & = & \frac{2}{N} \left[ \frac{1- m }{2} w \Bigl( 2 \beta u' _{\mathrm{eff} } \left( m \right) \Bigr) \right. \nonumber \\
  & & - \left. \frac{1+ m }{2} w \Bigl( -2 \beta u' _{\mathrm{eff} } \left( m \right) \Bigr) \right]  \nonumber \\
 & = & \frac{1}{N} \left\{ -m + \tanh \Bigl( -\beta u' _{\mathrm{eff} } (m) \Bigr) \right\} .
\label{dmdt1}
\end{eqnarray} 
Using the assumption that $N$ is large and $m$ can be approximated as $m \simeq \left< m \right>$, this equation can be transformed as follows: 
\begin{equation}
\frac{d m }{dt} = -m + \tanh \Bigl( -\beta u' _{\mathrm{eff} } (m) \Bigr) .
\label{dmdt2}
\end{equation} 
Comparing (\ref{dmdt2}) with (\ref{dmdt0}), we can evaluate $u_{\mathrm{eff} } (m)$ as follows:
\begin{eqnarray}
u_{\mathrm{eff} } (m) & = & -T \int dm \cdot \mathrm{tanh} ^{-1} \left( I_1 (m) \right) .
\label{ueff0}
\end{eqnarray} 

To calculate $I_1$ in Eq.~(\ref{ueff0}), we should first evaluate $p_m (x)$. In the steady state, the Fokker-Planck equation corresponding to Eq.~(\ref{Langevin2}) with a fixed $m$ is given as follows:
\begin{equation}
0 = -\frac{\partial}{\partial x} \left[ ( f_{\mathrm{ex} } + J'(x) m^2) p_m \right] + \frac{T}{N} \frac{\partial ^2 p_m}{\partial x^2} .
\label{FP1}
\end{equation} 
The solution of this equation is given as follows:
\begin{eqnarray}
p_m (x) & = &  e^{N \beta ( f_{\mathrm{ex} } x + J(x) m^2 ) } \nonumber \\
& & \cdot \left[ A + B \int _{x} ^{x_0} e^{-N \beta ( f_{\mathrm{ex} } \xi + J(\xi ) m^2 ) } d \xi \right] \nonumber \\
& = & e^{-N \beta u_{\mathrm{loc} } (x,m) }  \left[ A + B \int _{x} ^{x_0} e^{N \beta u_{\mathrm{loc} } (\xi ,m ) } d \xi \right] , \nonumber \\ 
& &
\label{FP_solution}
\end{eqnarray}  
where $A, B$, and $x_0$ are constants, and $u_{\mathrm{loc} }$ is defined as follows: 
\begin{equation}
u_{\mathrm{loc} } (x,m) \equiv -f_{\mathrm{ex} } x - J(x) m^2 .
\label{u_loc}
\end{equation} 
Using the value of $p_m (x_0)$, Eq.~(\ref{FP_solution}) can also be expressed as follows:
\begin{eqnarray}
p_m (x) & = & e^{-N \beta u_{\mathrm{loc} } (x,m) } \biggl[  e^{N \beta u_{\mathrm{loc} } (x_0 ,m) } p_m(x_0) \biggr. \nonumber \\
& & \left. + B \int _{x} ^{x_0 } e^{N \beta u_{\mathrm{loc} } (\xi ,m ) } d \xi \right] .
\label{FP_solution2}
\end{eqnarray} 
In particular, considering that $ e^{N \beta u_{\mathrm{loc} } (x_0 , m) } $ converges to zero as $x_0$ becomes sufficiently large, Eq.~(\ref{FP_solution2}) can be transformed as follows:
\begin{eqnarray}
p_m (x) & = & B e^{-N \beta u_{\mathrm{loc} } (x,m) } \int _{x} ^{\infty } e^{N \beta u_{\mathrm{loc} } (\xi ,m ) } d \xi .
\label{FP_solution3}
\end{eqnarray} 
In the case of the type-A model, the integral of Eq.~(\ref{FP_solution}) can be calculated because $u_{\mathrm{loc} } (\xi ,m)$ is a piecewise linear function of $\xi$. We will discuss the concrete form of $p_m$ for this case in appendix \ref{AppB}. Conversely, in the general cases, including the type-B model, the exact calculation of this integral is difficult. Hence, we evaluate the approximate value. 

First, we consider a case where $u_{\mathrm{loc} } $ has a local minimum, $x_m$. In this case, the system is thought to be in the local equilibrium state near this point. Hence, this case is thought to correspond to the stick state. When $u_{\mathrm{loc} }$ has a continuous derivative, $x_m$ is determined by the following equation:
\begin{equation}
\left. \frac{\partial u_{\mathrm{loc} } }{\partial x} \right| _{x=x_m} = - J'(x_m ) m^2 - f_{\mathrm{ex} } = 0 .
\label{x_loc}
\end{equation} 
In the type-A model, on the other hand, the local minimum of $u_{\mathrm{loc} }$ appears at the discontinuous point of $J'(x)$, i.e., $x=0$. Hence, $x_m$ is given by the following:
\begin{equation}
x_m = \left\{ 
\begin{array}{cc}
0 & \text{ (type-A) } , \\
\arcsin \left( \frac{f_{\mathrm{ex} } }{ J_1 m^2 } \right)  & \text{ (type-B) } ,
\end{array}
\right.
\label{xm1}
\end{equation} 
and exists if and only if
\begin{equation}
|m| > m_{\mathrm{th} } \equiv \left\{ 
\begin{array}{cc}
\sqrt{ \frac{\pi f_{\mathrm{ex} } }{2J_1} } & \text{ (type-A) } . \\
\sqrt{ \frac{f_{\mathrm{ex} } }{J_1} } & \text{ (type-B) } .
\end{array}
\right.
\label{xm2}
\end{equation} 
If $N$ is sufficiently large, $x$ is located on $x_m$ with high probability:
\begin{equation}
p_m (x) \simeq \delta (x - x_m) .
\label{P_loc}
\end{equation} 
Indeed, if $x_m$ exists, the integral of Eq.~(\ref{FP_solution3}) has a nearly constant value in the neighborhood of $x_m$:
\begin{equation}
\int _{x} ^{\infty } e^{N \beta u_{\mathrm{loc} } (\xi ,m ) } d \xi \sim e^{N \beta u_{\mathrm{loc} }\left( X_m ,m \right)  } .
\label{FP_int1}
\end{equation} 
Here, $X_m$ is the smallest argument of the local maximum of $u_{\mathrm{loc} }$, which is larger than $x_m$. Hence, $p_m$ is proportional to $e^{-N \beta u_{\mathrm{loc} }(x,m) }$ when $x \simeq x_m$, and this function draws a sharp peak at $x=x_m$.
Substituting Eq.~(\ref{P_loc}) into Eqs.~(\ref{I1_def}) and (\ref{ueff0}), we obtain the following:
\begin{eqnarray}
u_{\mathrm{eff} } (m) & = & - \int dm \cdot 2J(x_m) m \nonumber \\
& = & u_{\mathrm{loc} } (x_m ,m) + \mathrm{const.} \ . \label{ueff_stick} 
\end{eqnarray} 
The second line of Eq.~(\ref{ueff_stick}) can be derived by the following equation:
\begin{eqnarray}
\left. \frac{d u_{\mathrm{loc} } }{dm} \right| _{x = x_m } & = & \left. \frac{\partial u_{\mathrm{loc} } }{\partial m} \right| _{x = x_m } + \left. \frac{\partial u_{\mathrm{loc} } }{\partial x} \right| _{x = x_m } \cdot \frac{d x_m}{dm} \nonumber \\ 
& = &  \left. \frac{\partial u_{\mathrm{loc} } }{\partial m} \right| _{x = x_m } = -2J (x_m) m . \label{del_uloc} 
\end{eqnarray} 
Eq.(\ref{del_uloc}) is certified by Eq.~(\ref{x_loc}) in the case of the type-B model, and by the fact that $x_m$ is constant in the case of the type-A model. 

In the slip state where $|m| < m_{\mathrm{th} }$, the integrand of Eq.~(\ref{FP_solution3}) decreases monotonically and rapidly. Hence, the integral can be evaluated as follows:
\begin{eqnarray}
& & \int _{x} ^{x_0} e^{N \beta u_{\mathrm{loc} } (\xi ,m ) } d \xi \nonumber \\
& \simeq & e^{N \beta u_{\mathrm{loc} } (x,m) }  \int _{x} ^{\infty} e^{N \beta \partial _x u_{\mathrm{loc} } (x,m) \cdot (\xi -x) } d \xi \nonumber \\
& = & - \frac{e^{N \beta u_{\mathrm{loc} } (x,m) } }{N \beta \partial _x u_{\mathrm{loc} } (x,m) } .
\label{FP_Aint}
\end{eqnarray} 
Substituting Eq.~(\ref{FP_Aint}) into Eq.~(\ref{FP_solution3}), we obtain the following:
\begin{equation}
p_m (x) = - \frac{B' }{ \partial _x u _{\mathrm{loc} } (x,m) } = \frac{B' }{ f_{\mathrm{ex} } + J'(x) m^2 } .
\label{FP_A1}
\end{equation} 
Here, constant $B' \equiv B/(N\beta)$ is determined by the normalization condition. This equation indicates that the duration of stay at each $x$ is inversely proportional to velocity $v \sim (f_{\mathrm{ex} } + J'(x) m^2)/\gamma$. 

In short, in the case of the type-B model, we use Eq.~(\ref{ueff_stick}) in the stick state, and evaluate $I_1$  using Eq.~(\ref{FP_A1}) in the slip state. This evaluation can also be applied to the type-A model. However, as already explained, we do not adopt it for the type-A model because the integral of Eq.~(\ref{FP_solution}) is calculated in appendix \ref{AppB}.

\begin{figure}[hbp!]
\begin{center}
\includegraphics[width = 8.0cm]{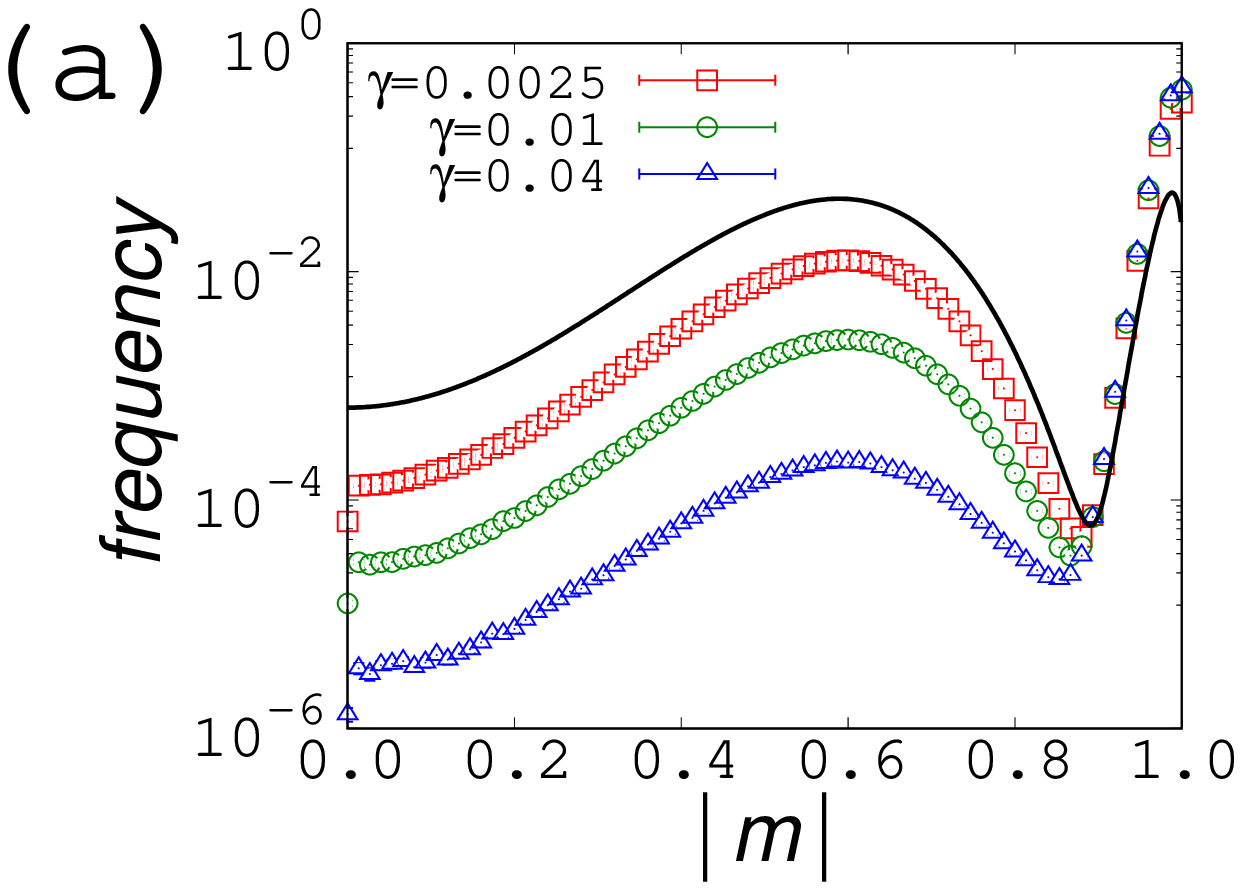} 
\includegraphics[width = 8.0cm]{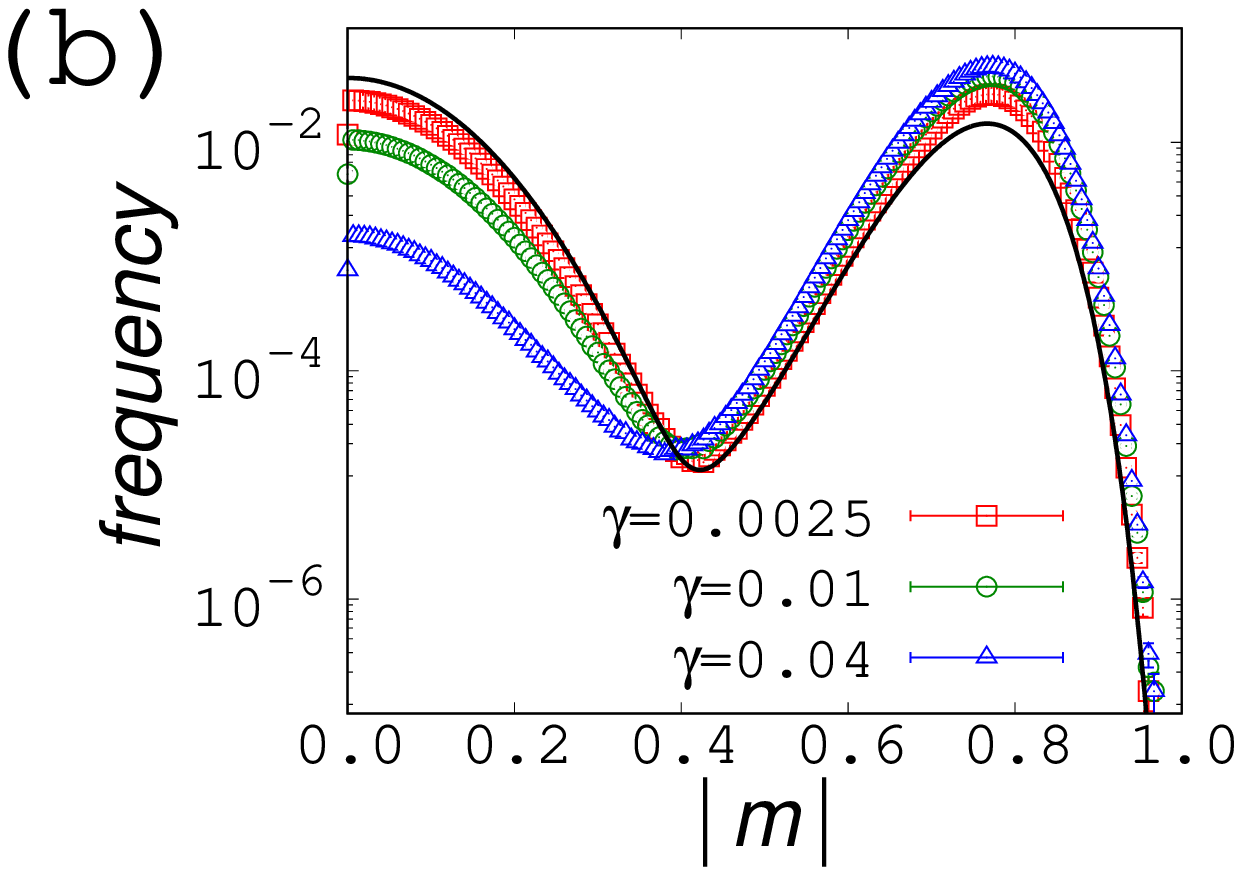} 
\end{center}
\caption{ Histograms of $m$ of the type-A model with different $\gamma$ values at (a) $T=1.5, N=150$, and  $f_{\mathrm{ex} }=0.5$ and (b) $T=3, N=300$, and $ f_{\mathrm{ex} }=0.1$. The red square, green circular, and blue triangular points represent the results of simulations at $\gamma = $0.0025, 0.01, 0.04, respectively. The black curves show the theoretical evaluation using Eq.~(\ref{pdf_eff}). }
\label{Histm_A}
\end{figure}

Using $a_{\mathrm{eff} } (m) $, the probability density of $m$, $P(m)$, is expressed as follows:
\begin{equation}
P(m) \propto e^{-\beta N a_{\mathrm{eff} } (m) } .
\label{pdf_eff}
\end{equation} 
We compare the result of Eq.~(\ref{pdf_eff}) and the numerical simulation to investigate the validity of the above discussions. The results of the type-A model are shown in Fig. \ref{Histm_A}. Note that in the histograms of this figure, we did not distinguish the value of $x$. Hence, these histograms are the integration of those calculated in Fig.~\ref{Histmx_A} over $x$, except for the fact that the parameters have different values. As seen in Fig.~\ref{Histm_A}, the results of the simulation seem to approach the estimation of Eq.~(\ref{pdf_eff}) with decreasing $\gamma$.

 Comparing Figs.~\ref{Histm_A} (a) and (b), the point $m=0$ changes from the argument of the local minimum of $P(m)$ to that of the local maximum with increasing $T$. It results from the change of the sign of $a _{\mathrm{eff} } '' (m)$ at $m=0$. Using Eqs.~(\ref{Free_slip}), (\ref{entropy1}), and (\ref{ueff0}), this quantity is calculated as follows:
\begin{eqnarray}
a _{\mathrm{eff} } '' (0) & = & u _{\mathrm{eff} } '' (0) - T s '' (0) \nonumber \\
& = & -T \frac{I ' _1(0) }{1 - I _1 ^2 (0) } + T
\label{dif2aeff_zero0}
\end{eqnarray} 
According to Eq.~(\ref{I1_def}), $I_1(0)$ is given as $I_1 (0) = 0$. Calculation of $I ' _1 (0)$ is explained in the appendix \ref{AppC}. Using the result of this appendix, Eq.~(\ref{dif2aeff_zero0}) can be transformed as follows:
\begin{equation}
a _{\mathrm{eff} } '' (0) =  -T \cdot (2 \beta J_0) + T = T - 2J_0 .
\label{dif2aeff_zero1}
\end{equation} 
Hence, when $T < 2 J_0( = 2)$, $a _{\mathrm{eff} } '' (0) $ is negative and $P(0)$ becomes the local minimum. In contrast, it becomes the local maximum when $T > 2 J_0( = 2)$.

\begin{figure}[hbp!]
\begin{center}
\includegraphics[width = 8.0cm]{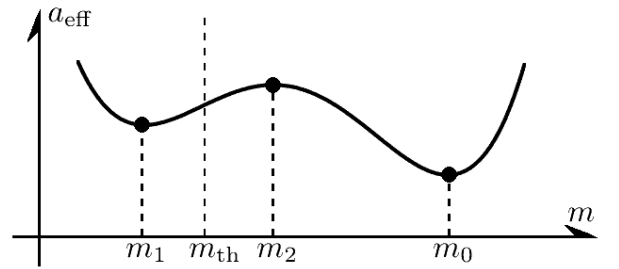} 
\end{center}
\caption{ Definitions of $m_0$, $m_1$, and $m_2$. As explained in Sec.~\ref{evaluation}, $m_0$ and $m_1$ are defined as the arguments of the local minima of $a_{\mathrm{eff} } (m) $ in the areas where $m > m_{\mathrm{th} } $ and $m < m_{\mathrm{th} } $, respectively. In Sec.~\ref{highg}, we also introduce the value $m_2$, which is the argument of the local maximum in the area where $ m_{\mathrm{th} } < m < m_0 $. }
\label{m0m1m2}
\end{figure}

\begin{figure}[hbp!]
\begin{center}
\includegraphics[width = 8.0cm]{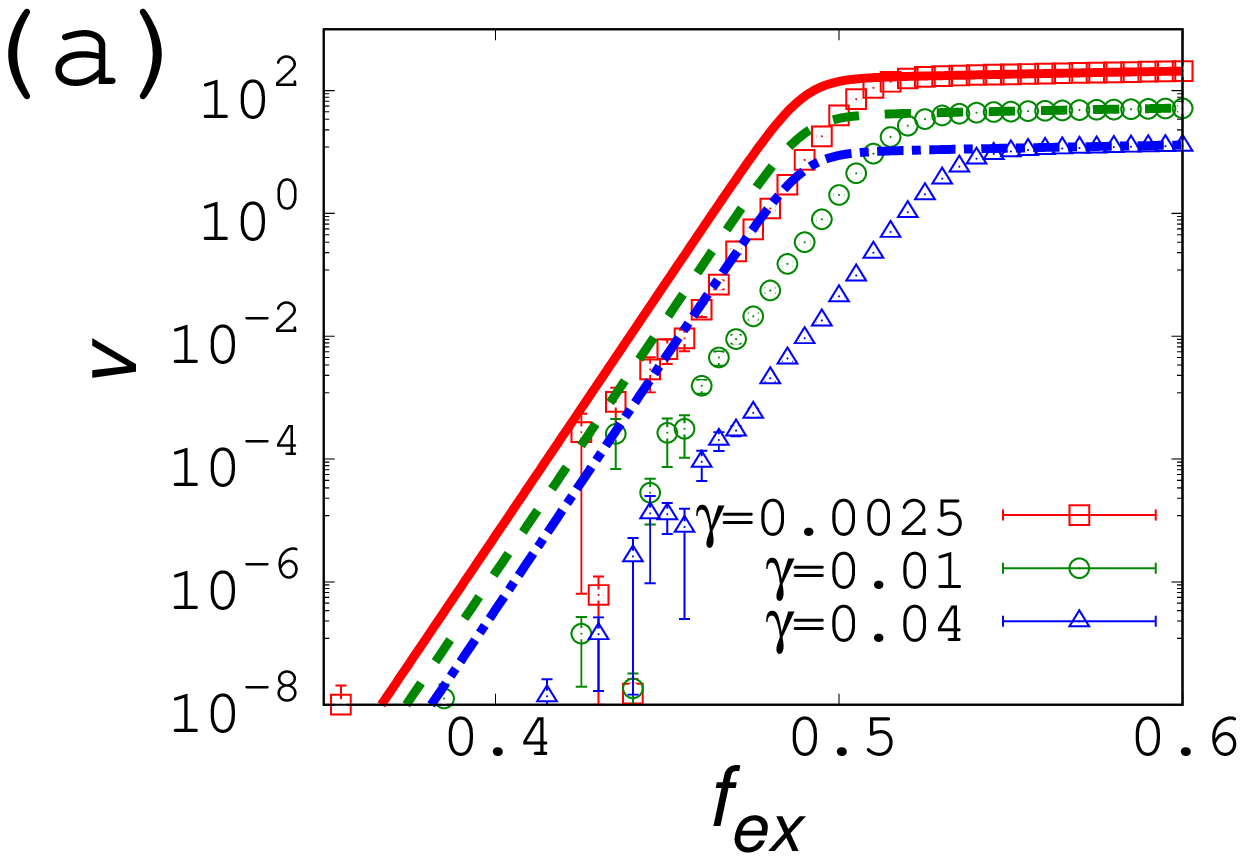} 
\includegraphics[width = 8.0cm]{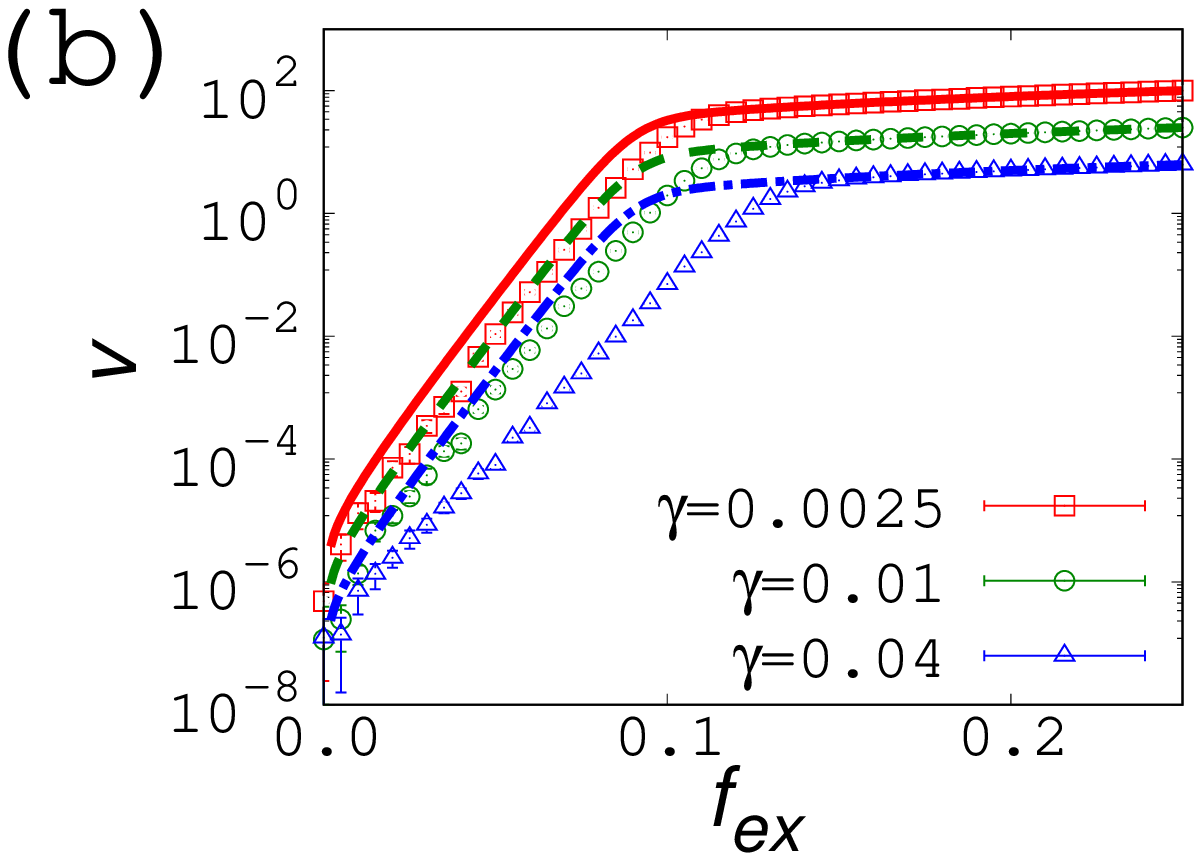} 
\end{center}
\caption{ $v$-$f_{\mathrm{ex} }$ relation of the type-A model with different $\gamma$ values at (a)$T=1.5$ and $N=150$, and (b)$T=3$ and $N=300$. The red square, green circular, and blue triangular points represent the results of simulations at $\gamma = $0.0025, 0.01, 0.04, respectively. The red solid, green dashed, and blue dash-dotted curves show the results of the theoretical evaluation using Eq.~(\ref{v_ave}) for each value of $\gamma$. }
\label{fv_A}
\end{figure}

We let $m_0$ and $m_1$ be the arguments of the local minima of $a_{\mathrm{eff} } (m) $ in the areas where $m > m_{\mathrm{th} } $ and $m < m_{\mathrm{th} } $, respectively, as shown in Fig.~\ref{m0m1m2}. 
Then, defining $q_0$ and $q_1$ as follows:
\begin{eqnarray}
q_0 & \equiv & e^{-\beta N a_{\mathrm{eff} } (m_0 )  } \int _{-\infty} ^{\infty} e^{ -\beta N a_{\mathrm{eff} } '' (m_0 ) \cdot (m-m_0 ) ^2 } d m \nonumber \\
& \propto & \frac{ e^{-\beta N a_{\mathrm{eff} } (m_0 )  } }{ \sqrt{ a_{\mathrm{eff} } '' (m_0 ) } } , \label{q0} \\
q_1 & \equiv & e^{-\beta N a_{\mathrm{eff} } (m_1 )  } \int _{-\infty} ^{\infty} e^{ -\beta N a_{\mathrm{eff} } '' (m_1 ) \cdot (m-m_1 ) ^2 } d m \nonumber \\
& \propto & \frac{ e^{-\beta N a_{\mathrm{eff} } (m_1 )  } }{ \sqrt{ a_{\mathrm{eff} } '' (m_1 ) } } , \label{q1}
\end{eqnarray} 
 the probabilities with which the stick and slip states appear are nearly proportional to $q_0$ and $q_1$, respectively. Hence, when both of these local minima exist, the slip state appears with probability
\begin{equation}
P_{\mathrm{slip} } = \frac{q_1  }{q_0 + q_1 }.
\label{P_slip}
\end{equation} 
Note that $P_{\mathrm{slip} } = 0$ when $m_1$ does not exist, and $P_{\mathrm{slip} } = 1$ when $m_0$ does not exist. The average velocity of the slip state is given by the following: 
\begin{equation}
\frac{ \int _{x=0} ^{x=2\pi} \frac{dx}{dt} dt }{ \int _{x=0} ^{x=2\pi} dt } = \frac{ 2 \pi }{ \int _{x=0} ^{x=2\pi} \left( \frac{dx}{dt} \right) ^{-1} dx } \simeq \frac{2 \pi}{\gamma I_2 (m_1) } ,
\label{v_slip}
\end{equation} 
\begin{equation}
\mathrm{where} \ \ I_2 (m) \equiv \int _{x=0} ^{x=2\pi} \left( f_{\mathrm{ex} }  + J'(x) m ^2 \right) ^{-1} dx .
\label{I2}
\end{equation} 
Using Eq.~(\ref{v_slip}), we obtain the expectation value of the velocity of this system:
\begin{equation}
v = \frac{2 \pi}{\gamma I_2 (m_1) } \cdot P_{\mathrm{slip} } .
\label{v_ave}
\end{equation} 
Note that although the velocity of the slip state given by Eq.~(\ref{v_slip}) itself has a large value under a small $\gamma$, the ensemble average of the velocity given by Eq.~(\ref{v_ave}) is small when the slip state rarely appears.
We compare the results of Eq.~(\ref{v_ave}) with those of the numerical simulation. The results of the type-A model are shown in Fig.~\ref{fv_A}. As seen in this figure, the estimation of Eq.~(\ref{v_ave}) seems to become accurate with decreasing $\gamma$, as is the case with the histogram of Fig.~\ref{Histm_A}. 

\begin{figure}[hbp!]
\begin{center}
\includegraphics[width = 8.0cm]{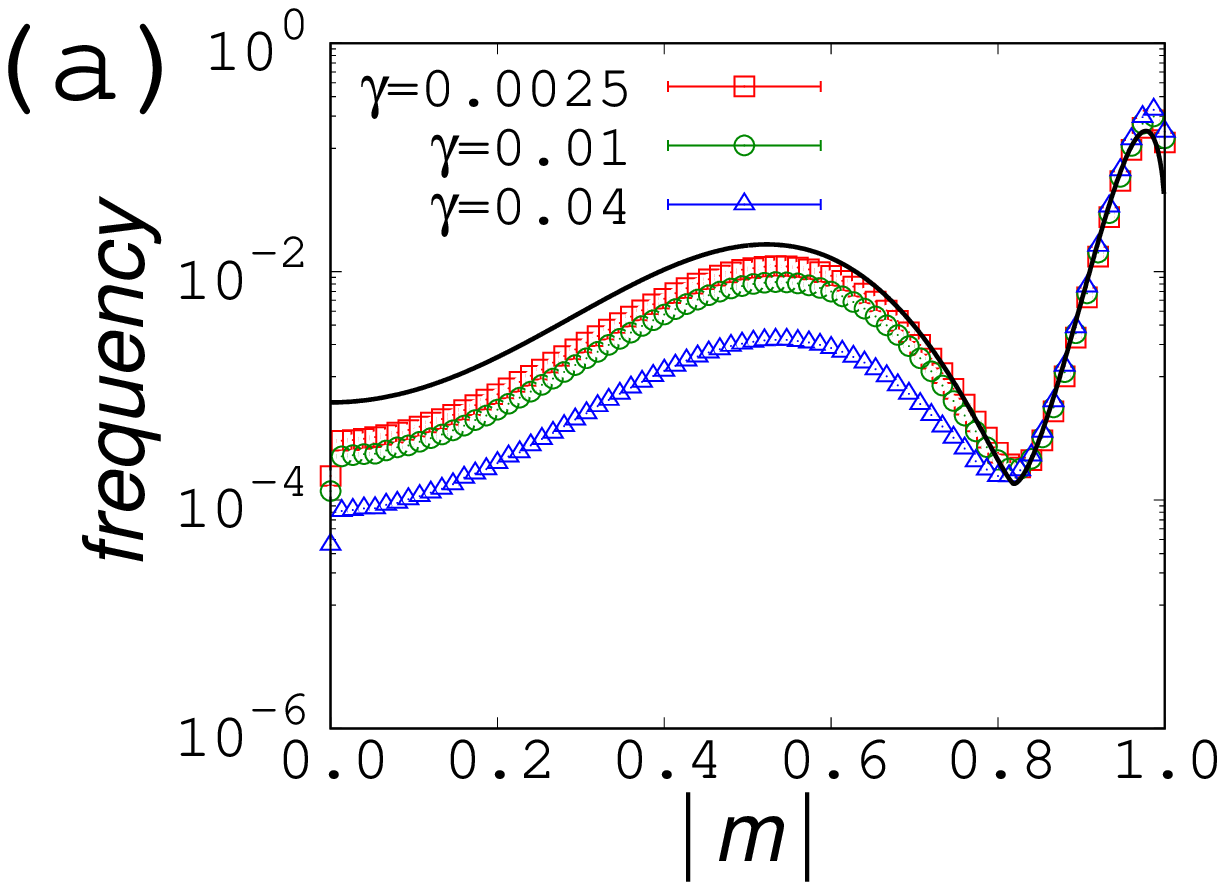}
\includegraphics[width = 8.0cm]{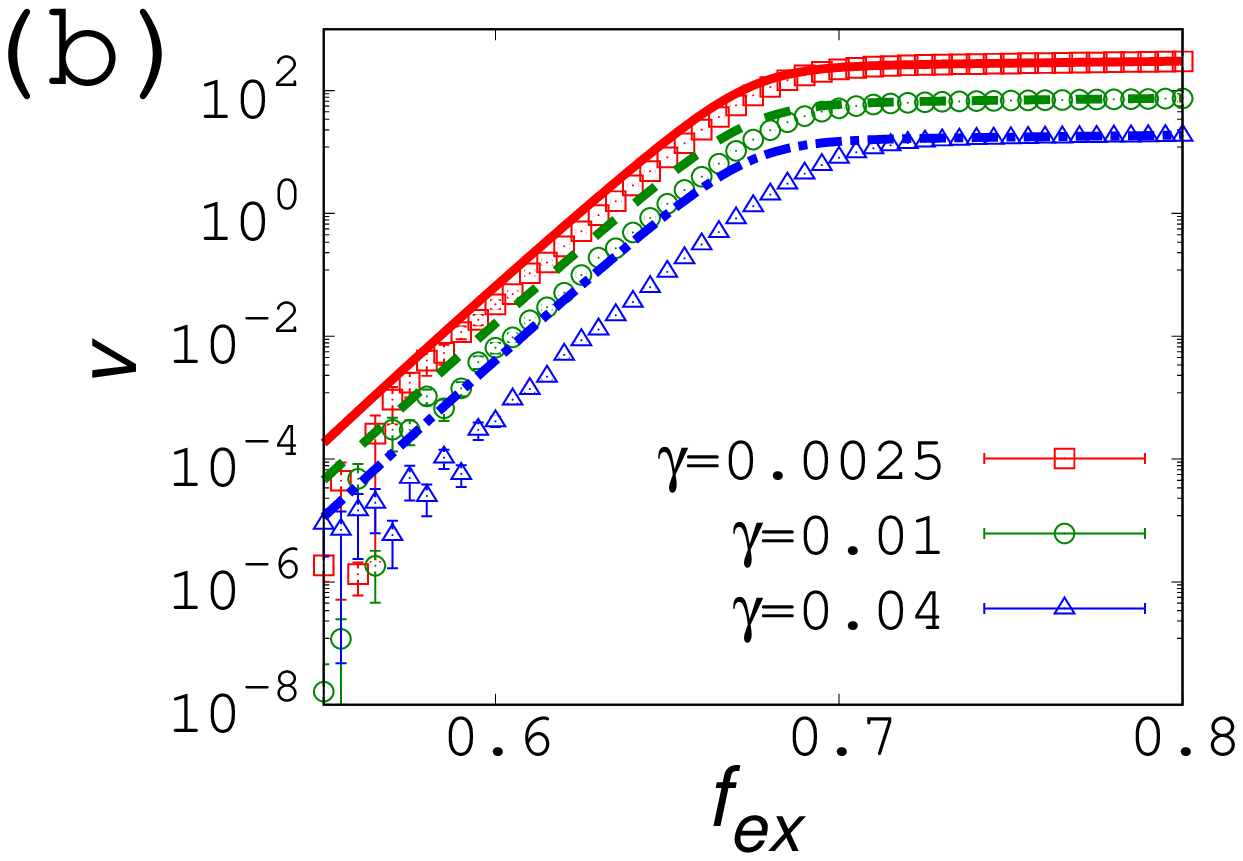} 
\end{center}
\caption{ $v$-$f_{\mathrm{ex} }$ relation of the type-B model with different $\gamma$ values at $T=1.5$ and $N=150$. The meanings of the points and curves are the same as those in Figs.~\ref{Histm_A} and \ref{fv_A}. }
\label{fv_B}
\end{figure}

Calculations similar to those shown in Figs.~\ref{Histm_A} and \ref{fv_A} in the case of the type-B model are shown in Fig.~\ref{fv_B}. As seen in this figure, the results of the simulations seem to converge to those of Eqs.~(\ref{pdf_eff}) and (\ref{v_ave}) with decreasing $\gamma$, as is the case with the type-A model. As already mentioned, in the case of the type-B model, we used Eqs.~(\ref{ueff_stick}) and (\ref{FP_A1}), the rough approximation of Eq.~(\ref{FP_solution}), to evaluate  $a_{\mathrm{eff} }$, whereas we used Eq.~(\ref{FP_solution}) itself for the type-A model. Hence the accuracy of the theoretical evaluation is thought to be higher for the type-A model than for the type-B model. However, a comparison of Figs.\ref{Histm_A}, \ref{fv_A}, and \ref{fv_B} does not show this tendency. It is difficult to infer the reason why such behavior occurs, but one possible factor is the point that the dependence of histograms on $\gamma$ is too large to observe such tendency. Another point to note is the form of $p_m$ for the type-A model under Eq.~(\ref{FP_A1}). In the case of the type-B model, both $p_m$ at $|m| = m_{\mathrm{th} }+0$ evaluated by Eq.~(\ref{P_loc}) and that at $|m| = m_{\mathrm{th} }-0$ evaluated by Eq.~(\ref{FP_A1}) have a sharp peak near $x = \pi/2$. On the other hand, in the type-A model, $p_m$ is calculated as a piecewise constant function at $|m| < m_{\mathrm{th} }$ if Eq.~(\ref{FP_A1}) is adopted. This evaluation has an apparently different form from  Eq.~(\ref{P_loc}). It is thought that such a difference makes the evaluation of $p_m$ and $a_{\mathrm{eff} } $ near $|m| = m_{\mathrm{th} }$ using Eqs.~(\ref{ueff_stick}) and (\ref{FP_A1}) inaccurate and the improved approximation is required for the type-A model.

\subsection{Comparison of cases with small and large values for $\gamma$ \label{highg} }

\begin{figure}[hbp!]
\begin{center}
\includegraphics[width = 8.0cm]{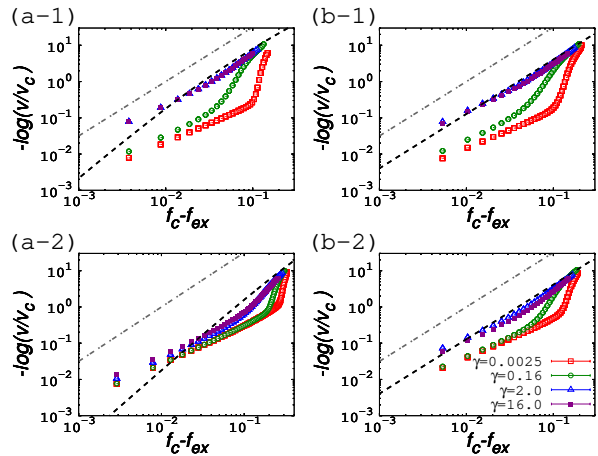} 
\end{center}
\caption{ Relation between $-\log (v/v_c)$ and $f_c - f_{\mathrm{ex} }$ in cases for (a-1)type-A model at $T=1.5$ and $N=150$, (a-2)type-A model at $T=3$ and $N=300$, (b-1)type-B model at $T=1.5$ and $N=150$, and (b-2)type-B model at $T=3$ and $N=300$. The red open square, green circular, blue triangular, and purple closed square points indicate the results of simulations at $\gamma = 0.0025$, 0.16, 2, 16, respectively, and the black dashed lines or curves were calculated using Eq.~(\ref{v_highg}). We also plotted the proportional relation of  Eq.~(\ref{PT}), as shown by the gray dash-dotted lines. }
\label{vf_highg}
\end{figure}

In a case where $\gamma$ is large, as we saw in Fig.~\ref{Histmx_A}, the coexistence of the stick and slip states as metastable states breaks down. We evaluate the velocity of this case, assuming that the lattice motion is caused by the thermal activation process. Namely, we first define $\Delta a_{\mathrm{eff} }$ as the difference between the local maximum and minimum in the region where $ m_{\mathrm{th} } < |m| \leq m_0 $:
\begin{equation}
\Delta a_{\mathrm{eff} } \equiv a_{\mathrm{eff} } (m_2) - a_{\mathrm{eff} } (m_0) ,
\label{del_aeff}
\end{equation} 
and consider that the system is trapped in the stick state by the free energy barrier given as $\Delta a_{\mathrm{eff} } $. Note that $m_0$ was already defined in Sec.~\ref{evaluation}, and $m_2$ is the argument of the local maximum in the area where $ m_{\mathrm{th} } < m \leq m_0 $, as shown in Fig.~\ref{m0m1m2}. Under this assumption, velocity $v$ obeys the following relation:
\begin{equation}
v \propto e^{ -\beta N \Delta a_{\mathrm{eff} }  } .
\label{v_highg}
\end{equation} 
If external force $f_{\mathrm{ex} }$ is stronger than a certain value, $f_c$, the magnetic structure cannot trap the lattice motion, and Eq.~(\ref{v_highg}) breaks down. The point $f_{\mathrm{ex} } = f_c$ is expressed as the critical value where the local maximum and minimum of $a_{\mathrm{eff} }$ disappear and $\Delta a_{\mathrm{eff} } $ cannot be defined. We calculate the relation between $-\log (v/v_c)$ and $f_c - f_{\mathrm{ex} } $ using Eq.~(\ref{v_highg}), and compare the results with those of the simulation. Here, $v_c$ is the value of $v$ at $f_{\mathrm{ex} } = f_c$. As the value of $f_c$, we use the result of the theoretical evaluation even in the case of the simulation. The results are shown in Fig.~\ref{vf_highg}. As seen in this figure, the $v$-$f_{\mathrm{ex} }$ relation shows apparently different behaviors depending on $\gamma$, and is close to the results of Eq.~(\ref{v_highg}) drawn as the black dashed lines (or curves,) when $\gamma$ is large. The gray dash-dotted lines of Fig.~\ref{vf_highg} indicate the proportional relation of  Eq.~(\ref{PT}). In the case of the type-B model, these two lines are parallel to each other, which means that the evaluation of $v$ using Eq.~(\ref{v_highg}) obeys Eq.~(\ref{PT}). This behavior results from the fact that Eq.~(\ref{v_highg}) premises the continuous change between the stick and slip states. The value of $\log v$ given by Eq.~(\ref{v_highg}) is not a linear function of $f_{\mathrm{ex} } $, because $m _0 - m_2$ decreases with decreasing $f_c - f_{\mathrm{ex} } $. Instead, $\log v$ actually obeys Eq.~(\ref{PT}) if $J(x)$ does not have singular points. This mechanism resembles to that of the Prandtl-Tomlinson model. (For a more detailed explanation, see Ref.~\cite{PAMPSS03}, for example.) In short, the lattice motion of our model in the large-$\gamma$ case is essentially the same as that of the Prandtl-Tomlinson model, and the instantaneous phase transition proposed by Ref.~\cite{PAMPSS03} does not exist. Note that in the case of the type-A model, which has a singular point of $J(x)$ at $x=0$, the result of Eq.~(\ref{v_highg}) does not obey Eq.~(\ref{PT}).

\begin{figure}[hbp!]
\begin{center}
\includegraphics[width = 8.0cm]{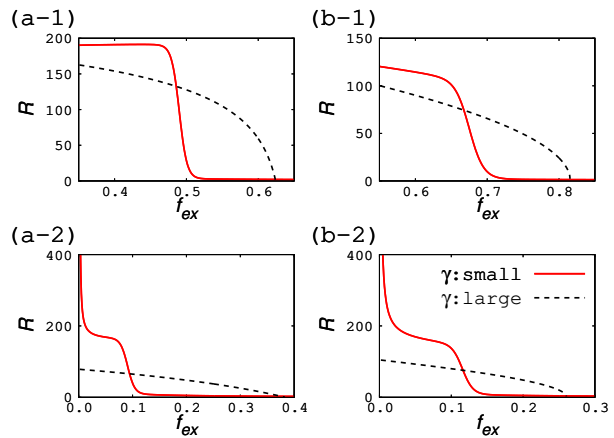} 
\end{center}
\caption{ $R$-$f_{\mathrm{ex} }$ relations of (a-1) type-A model at $T=1.5$ and $N=150$, (a-2) type-A model at $T=3$ and $N=300$, (b-1) type-B model at $T=1.5$ and $N=150$, and (b-2) type-B model at $T=3$ and $N=300$. The red solid curves are the results of Eq.~(\ref{v_ave}), and the black dashed curves are those of Eq.~(\ref{v_highg}). }
\label{Rv1}
\end{figure}

Finally, we calculate the quantity
\begin{equation}
R \equiv \frac{d}{d f_{\mathrm{ex} } } \left( \log v \right) ,
\label{R_def}
\end{equation} 
to investigate the difference compared to the Dieterich-Ruina law. This quantity becomes constant if the system obeys Eq.~(\ref{DR}). When $v$ is small, it is difficult to calculate $R$ using the numerical differentiation of the simulation results, because the error bars of $\log v$ are large. Hence, we instead use the results of approximate evaluations based on Eqs.~(\ref{v_ave}) and (\ref{v_highg}). The results of these two equations are compared in Fig.~\ref{Rv1}. Note that the values of $R$ calculated by these equations do not depend on $\gamma$.
As seen in Fig.~\ref{Rv1}, the results of Eq.~(\ref{v_ave}), which correspond to the small-$\gamma$ case, show a plateau in the small-$f_{\mathrm{ex} }$ range. In our previous studies, we considered that the system obeyed the Dieterich--Ruina law in this range. However, even in this plateau, the value of $R$ is not always constant. Hence, strictly speaking, the Dieterich--Ruina law does not always hold even in this range. It is possible that the models of our previous studies also deviated from the Dieterich--Ruina law, but we could not find the difference using only the $v$-$f_{\mathrm{ex} }$ graphs, like those shown in Figs.~\ref{fv_A} and \ref{fv_B}.(b). We briefly discuss this deviation using Eq.~(\ref{v_ave}). When $f_{\mathrm{ex} }$ is small, the stick state appears with high probability. It means that $q_0 \gg q_1$, and consequently Eq.~(\ref{v_ave}) gives the following relation:
\begin{equation}
v \propto \frac{q_0}{q_1} \propto e^{\beta N \left[ a_{\mathrm{eff} } (m_1) - a_{\mathrm{eff} } (m_0) \right]} .
\label{v_ave_pla}
\end{equation} 
Here, we ignored slowly varying factors. Using Eq.~(\ref{v_ave_pla}), $R$ is calculated as follows:
\begin{equation}
R = \beta N \frac{d}{d f_{\mathrm{ex} } } \left[ a_{\mathrm{eff} } (m_1) - a_{\mathrm{eff} } (m_0) \right] .
\label{R_pla}
\end{equation} 
Hence, in the small-$f_{\mathrm{ex} }$ range, $R$ stays constant and Eq.~(\ref{DR}) holds, if $a_{\mathrm{eff} } (m_0)$ and $a_{\mathrm{eff} } (m_1)$ is the linear function of $f_{\mathrm{ex} }$. In other words, deviation from Eq.~(\ref{DR}) in this range expresses the nonlinearities of the effective free energies, $a_{\mathrm{eff} } (m_0)$ and $a_{\mathrm{eff} } (m_1)$, as the function of $f_{\mathrm{ex} }$.
Note that in the large-$\gamma$ case, the discrepancy between Eq.~(\ref{DR}) and the actual velocity is caused by the similar mechanism to that of the Prandtl-Tomlinson model, as we already explained.


\section{Summary \label{summary} }

In this paper, we considered a simplified model of magnetic friction in which the lattice motion is prevented by the magnetization of the system. Numerical simulations showed that the mechanism of the lattice motion differed depending on the resistance coefficient, $\gamma$. When $\gamma$ is small, the stick and slip states exist as metastable states separated from each other, and the lattice velocity is determined by the probability of the slip state. In this case, the change between these two states resembles the first-order phase transition. As explained in Sec.~\ref{Histmx}, this transition has a different form from the inference of Ref.~\cite{PAMPSS03}, which did not consider metastability. On the other hand, when $\gamma$ is large, such a coexistence of two states is not observed, and the lattice motion is caused by the thermal activation process. This difference affects the relation between the lattice velocity and frictional force. As explained in Sec.~\ref{highg}, although the behavior of the lattice velocity at a small $\gamma$ value under a sufficiently weak external force seems to be close to the Dieterich--Ruina law, Eq.~(\ref{DR}), it does not always exactly coincide with this law.

Magnetic structures such as the magnetization are a kind of internal structure of the lattice. Hence, it is expected that the results of this study could be applied not only to magnetic friction, but also to the friction of normal solid surfaces. Such a generalization of this study could be the subject of future work. In particular, whether the stick and slip states are separated as metastable states or interchange with each other smoothly should also be investigated in the case of the usual solid surfaces. 
To apply our study to the realistic friction, simplification of the model seems to become the problem. For example, the overdamped Langevin equation assumes that the inertial term is much smaller than the damping term. Hence, our investigation on this equation under small damping constant $\gamma$ seems to be strange. However, it does not mean that our study failed to grasp the essential behavior of the friction. In our model, the velocity of the slip state is fast when $\gamma$ is small. Considering that the system does not have sufficient time to change the magnetization under the fast lattice motion, high velocity of the slip state is thought to be the direct cause of the separation of the stick and slip states. Hence, it is possible that such separation can be observed even though $\gamma$ itself does not have a small value. In addition, in the case of the system which does not have the parameter corresponding to $\gamma$, the velocity of the slip state is thought to be the alternative criterion which judges whether such separation exists. We should consider these points carefully in future work.

The system size $N$ should be also remarked. In this study, we considered the case that $N$ is several hundred. If $N$ gets larger, the relaxation time of the system diverges rapidly and the stochastic transition between two metastable states cannot be observed. On the other hand, if $N=1$, metastable states themselves do not appear because there are no internal structures such as $m$. Hence, the coexistence of the metastable stick and slip states discussed in our study can be observed only when $N$ is not too large nor too small. This restriction seems to be related with the true contact area of the solid surfaces. In the case of the friction of the normal solid surfaces, it is known that the frictional force is generated by the asperities, the junctions of protrusions of the surfaces. Hence, the true contact area of the surfaces is much smaller than the apparent one, but it is never composed of only one atom or molecule\cite{KHKBC12}. In short, both the system we investigated in this study and the true contact area of the solid surfaces are few-body systems, and our result is an example that friction can have the behavior peculiar to such systems. Considering this point, our result implies that the dependence on the size of the contact area has an important role on the friction.

From the standpoint of statistical mechanics, it is also interesting that the infinite-range Ising model, which shows the second-order phase transition in the equilibrium state, shows the first-order-like transition when it is combined with the lattice motion. In a case where the model has the first-order phase transition or does not have any phase transition in the equilibrium state, whether the system behaves the same as our present model remains unclear. This point should also be studied in future.

\acknowledgements

This study was supported by JSPS KAKENHI (Grant Number JP21K13857). 
The computation in this work was performed using the facilities of the Supercomputer Center, the Institute for Solid State Physics, the University of Tokyo. We would like to thank Editage (www.editage.com) for English language editing.

\appendix 

\section{The border $\gamma _c$ where the behavior of the system changes \label{AppA}}

\begin{figure}[hbp!]
\begin{center}
\includegraphics[width = 8.0cm]{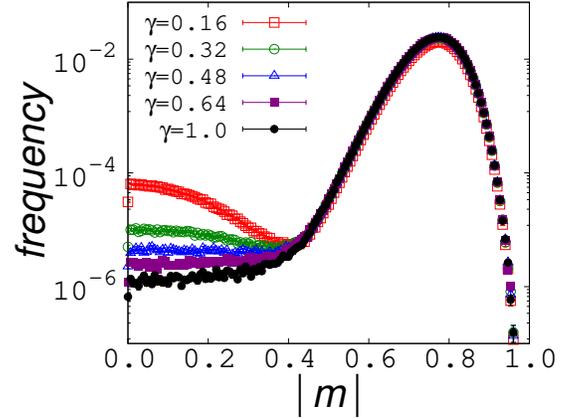} 
\end{center}
\caption{ Histogram of $m$ of the type-A model with different $\gamma$ values at $T=3, N=300$, and $ f_{\mathrm{ex} }=0.15$, under the condition that $0 \leq x < 0.02 \pi$. The red open square, green open circular, blue triangular, purple closed square, and black closed circular points represent the data at $\gamma = 0.16$, 0.32, 0.48, 0.64, 1.0, respectively. }
\label{Histm_x0_A}
\end{figure}

As we mentioned in the main text, the process of the lattice motion of our model shows two different behaviors depending on the value of $\gamma$. Namely, the stick and slip states are separated as the metastable states when $\gamma$ is small, and such separation is not observed when $\gamma$ is large. Finding the border $\gamma _c$ where the behavior changes is difficult because the value of $x$ keeps changing in the slip state, although this border is important. 
In this appendix, we introduce a trial to find it. Specifically, we calculate the histogram of $m$ when $x$ is near the $x_m$ given by Eq.~(\ref{xm1}), the point where the system under the stick state is most likely to be located. Hence, this histogram is the cross section of Fig. \ref{Histmx_A}, except for the fact that the parameters have different values. Note that this histogram is different from Fig. \ref{Histm_A} which shows the integration of Fig. \ref{Histmx_A}. 
If this histogram shows two peaks, we judge that there is a route for the slip state that keeps the lattice moving without changing into the stick state, and hence the separated metastable states exist. If there is only one peak, on the other hand, we think that the system should become the stick state when $x \simeq x_m$, and such separation does not exist. 
Fig. \ref{Histm_x0_A} shows the example of this histogram at $T=3, N=300$, and $ f_{\mathrm{ex} }=0.15$. Here, we take the data under the condition that $0 \leq x < 0.02 \pi$, i.e., $x \simeq x_m =0$. Seeing this figure, we can confirm that the second peak disappears in the range $0.48 \lesssim \gamma \lesssim 0.64 $, and the border $\gamma _c$ falls within this range. To calculate the exact value of $\gamma _c$, more careful simulation is required. 
In addition, $\gamma _c$ changes its value depending on $f_{\mathrm{ex}}$. The computational cost required to draw the $\gamma _c$-$f_{\mathrm{ex}}$ curve is thought to be large.

\section{Calculation of $p_m$ for the type-A model \label{AppB} }

In the case of the type-A model, using constants $a_1, a_2, b_1$, and $b_2$, Eq.~(\ref{FP_solution}) is expressed as follows:
\begin{equation}
p_m (x) = \left\{ 
\begin{array}{cc}
a_1 \exp \left( N \beta \alpha _{-,m} (x-\pi) \right)  + b_1 & (0 \leq x < \pi) , \\
a_2 \exp \left( N \beta \alpha _{+,m} x \right)  + b_2 & (-\pi \leq x < 0) ,
\end{array}
\right.
\label{pm_B1}
\end{equation} 
\begin{equation}
\mathrm{where} \ \ \alpha _{\pm ,m} \equiv f_{\mathrm{ex} } \pm \frac{2 J_1}{\pi} m^2.
\label{pm_B_alpha}
\end{equation} 
When $N\beta \gg 1$ and $|m|$ is sufficiently smaller than $m_{\mathrm{th} } $, $p_m$ is inversely proportional to the average velocity, $(f_{\mathrm{ex}} + J'(x) m^2)/\gamma $, as in the case of the type-B model (see Eq.~(\ref{FP_A1})). Hence, $b_1$ and $b_2$ can be expressed as follows:
\begin{equation}
b_1 = \frac{a}{\alpha _{-,m} } , \ \ b_2 = \frac{a}{\alpha _{+,m} } ,
\label{pm_B_b}
\end{equation} 
using constant $a$. The continuity of $p_m$ at $x=0, \pi$ gives the following relation:
\begin{subequations}
\begin{eqnarray}
& & a_1 e^{-N \beta \alpha _{-,m} \pi} +  \frac{a}{\alpha _{-,m} } = a_2 + \frac{a}{\alpha _{+,m} } , \label{pm_B_c1} \\
& & a_1 + \frac{a}{\alpha _{-,m} } =  a_2 e^{-N \beta \alpha _{+,m} \pi} + \frac{a}{\alpha _{+,m} } . \label{pm_B_c2}
\end{eqnarray} 
\end{subequations}
Solving these equations, we obtain relations between constants $a_1, a_2$, and $a$. When $N$ is large, $e^{-N \beta \alpha _{+,m} \pi}$ is much smaller than $e^{-N \beta \alpha _{-,m} \pi}$. Hence, we ignore the first term on the right-hand side of Eq.~(\ref{pm_B_c2}). Under this approximation, Eq.~(\ref{pm_B1}) can be rewritten as follows:
\begin{widetext}
\begin{equation}
p_m (x) = \left\{ 
\begin{array}{cc}
a \left( \frac{1}{\alpha _{+,m} } - \frac{1}{\alpha _{-,m} } \right) \exp \left( N \beta \alpha _{-,m} (x - \pi) \right)  + \frac{a}{\alpha _{-,m} } & (0 \leq x < \pi) , \\
- a\left( \frac{1}{\alpha _{+,m} } - \frac{1}{\alpha _{-,m} } \right) \left( 1 - e^{-N \beta \alpha _{-,m} \pi } \right) \exp \left( N \beta \alpha _{+,m} x \right)  + \frac{a}{\alpha _{+,m} } & (-\pi \leq x < 0) .
\end{array}
\right.
\label{pm_B2}
\end{equation} 
\end{widetext}
Letting $a' \equiv a e^{-N \beta \alpha \alpha _{-,m} \pi}$, Eq.~(\ref{pm_B2}) can be rewritten as follows:
\begin{widetext}
\begin{equation}
p_m (x) = \left\{ 
\begin{array}{cc}
a' \left( \frac{1}{\alpha _{+,m} } - \frac{1}{\alpha _{-,m} } \right) \exp \left( N \beta \alpha _{-,m} x \right)  + \frac{a'}{\alpha _{-,m} } e^{N \beta \alpha _{-,m} \pi} & (0 \leq x < \pi) , \\
a' \left( \frac{1}{\alpha _{+,m} } - \frac{1}{\alpha _{-,m} } \right) \left( 1 - e^{N \beta \alpha _{-,m} \pi } \right) \exp \left( N \beta \alpha _{+,m} x \right)  + \frac{a'}{\alpha _{+,m} } e^{N \beta \alpha _{-,m} \pi} & (-\pi \leq x < 0) .
\end{array}
\right.
\label{pm_B3}
\end{equation} 
\end{widetext}
This expression is convenient when $|m| > m_{\mathrm{th} }$. Note that $p_m$ given by Eq.(\ref{pm_B3}) has a sharp peak at $x = 0$ when $|m| > m_{\mathrm{th} }$ and $N$ is sufficiently large. Hence, this equation is not contradictory to Eq.~(\ref{P_loc}), although Eq.~(\ref{pm_B_b}) was derived from Eq.~(\ref{FP_A1}). In a case where $|m| = m_{\mathrm{th} }$, which means that $\alpha _{-,m} = 0$, the form of $p_m$ is expressed by taking the limit of Eq.(\ref{pm_B3}):  
\begin{widetext}
\begin{equation}
p_m (x) = \left\{ 
\begin{array}{cc}
a'' \left( \frac{1}{N \beta\alpha _{+,m} } + \pi - x \right)  & (0 \leq x < \pi) , \\
a'' \left( \frac{1 }{N \beta \alpha _{+,m} } + \pi \exp \left( N \beta \alpha _{+,m} x \right) \right) & (-\pi \leq x < 0) ,
\end{array}
\right.
\label{pm_B4}
\end{equation} 
\end{widetext}
where $a'' \equiv N \beta a'$.

\section{Calculation of $I' _1(0)$ \label{AppC} }

In this appendix, the value of $I' _1 (0)$ is calculated. First, we use the following relations:
\begin{equation}
\tanh \bigl( 2\beta J(x) m \bigr) = 2\beta J(x) m  + O \left( m^2 \right) ,
\label{tanh_nearzero}
\end{equation} 
\begin{eqnarray}
p_m (x) & = & B e^{N \beta f_{\mathrm{ex} } x } \int _{x} ^{\infty } e^{-N \beta f_{\mathrm{ex} } x } d \xi + O \left( m^2 \right) \nonumber \\ 
& = & \mathrm{const. } + O \left( m^2 \right) .
\label{pm_nearzero}
\end{eqnarray} 
Here, Eq.~(\ref{pm_nearzero}) is derived by using Eq.~(\ref{FP_solution3}). Substituting Eqs.~(\ref{tanh_nearzero}) and (\ref{pm_nearzero}) into Eq.~(\ref{I1_def}), $I(m)$ can be evaluated as follows:
\begin{eqnarray}
I (m) & = & \frac{ \int _0 ^{2\pi} 2\beta J(x) m dx }{ \int _0 ^{2\pi} dx } + O \left( m^2 \right) \nonumber \\ 
& = & 2\beta m \cdot \frac{ \int _0 ^{2\pi} J(x) dx }{2\pi } + O \left( m^2 \right) .
\label{I_nearzero0}
\end{eqnarray} 
In both type-A and B models, the integral appeared in the right-hand side of Eq.~(\ref{I_nearzero0}) is $2 \pi J_0$. Hence, this equation can be transformed as follows:
\begin{equation}
I(m) = 2\beta J_0 m  + O \left( m^2 \right) ,
\label{I_nearzero1}
\end{equation} 
and we obtain the value of $I'(0)$:
\begin{equation}
I' (0) = 2\beta J_0 .
\label{difI_zero}
\end{equation}

\end{document}